\documentclass[aps,groupedaddress]{revtex4b3}
\usepackage[dvips]{graphicx}
\begin{document}

\title{Tunneling conductance of SIN junctions with different gap
symmetries \\ and non-magnetic impurities \\by direct solution of
real-axis Eliashberg equations}

\author{G.A.\@ Ummarino}
\email[Corresponding author. E-mail:~]{ummarino@polito.it}
\affiliation{{\it INFM-}Dipartimento di Fisica, Politecnico di
Torino, Corso Duca degli Abruzzi, 24 - 10129 Torino, Italy}
\author{R.S.\@ Gonnelli}
\affiliation{{\it INFM-}Dipartimento di Fisica, Politecnico di
Torino, Corso Duca degli Abruzzi, 24 - 10129 Torino, Italy}
\author{D.\@ Daghero}
\affiliation{{\it INFM-}Dipartimento di Fisica, Politecnico di
Torino, Corso Duca degli Abruzzi, 24 - 10129 Torino, Italy}

\date{\today}

\newcommand{\ohm}{\ensuremath{{\rm\Omega}}}
\newcommand{\ped}[1]{\ensuremath{_{\rm #1}}}
\newcommand{\unit}[1]{\ensuremath{{\rm\,#1}}}
\newcommand{\gei}{\ensuremath{{\rm j}}}
\newcommand{\eu}{\ensuremath{{\rm e}}}
\newcommand{\micro}{\ensuremath{\mu}}

\begin{abstract}
We theoretically investigate the effect of various symmetries of
the superconducting order parameter $\Delta (\omega )$ on the
normalized tunneling conductance of SIN junctions by directly
solving the real-axis Eliashberg equations (EEs) for a half-filled
infinite band, with the simplifying assumption $\mu ^{\ast
}$~=~0. We analyze six different symmetries of the order parameter: $s$, $d$, $%
s+{\rm i}d$, $s+d$, {\it extended}~$s$ and {\it anisotropic}~$s$,
by assuming that the spectral function $\alpha ^{2}F(\Omega )$
contains an isotropic part $\alpha^{2}F(\Omega )_{{\rm is}}$ and
an anisotropic one, $\alpha^{2}F(\Omega )_{{\rm an}}$, such that
$\alpha ^{2}F(\Omega)_{{\rm an}}$=$g\cdot \alpha^{2}F
(\Omega)_{{\rm is}}$, where $g$ is a constant.

We compare the resulting conductance curves at $T=2$~K to those
obtained by analytical continuation of the imaginary-axis solution
of the EEs, and we show that the agreement is not equally good for
all symmetries. Then, we discuss the effect of non-magnetic
impurities on the theoretical tunneling conductance curves at
$T=4$~K for all the symmetries considered.

Finally, as an example, we apply our calculations to the case of
optimally-doped high-$T_{\mathrm{c}}$ superconductors.
Surprisingly, although the possibility of explaining the very
complex phenomenology of HTSC is probably beyond the limits of the
Eliashberg theory, the comparison of the theoretical curves
calculated at $T$=4~K with the experimental ones obtained in
various optimally-doped copper-oxides gives fairly good results.
\end{abstract}


\maketitle



\section{Introduction}
The semi-phenomenological Migdal-Eliashberg
theory~\cite{Eliashberg} has been successfully used in the past to
describe many properties of low-$T\ped{c}$ superconductors. For
example, it has provided a quite precise explanation of the
tunneling experimental data obtained on almost all the
conventional superconductors. In most of these cases, theoretical
predictions in good agreement with the experimental results have
been obtained by solving the Eliashberg equations (EEs) for an
$s$-wave order parameter. However, other symmetries of the order
parameter may exist. It is therefore interesting to study how, in
the framework of this theory, the gap symmetry influences the
theoretical conductance curves.

In this paper, we will calculate the theoretical normalized
conductance curves of SIN junctions for various gap symmetries
($s$, $d$, $s+{\rm i}d$, $s+d$, {\it anisotropic}~$s$ and {\it
extended}~$s$)~\cite{VanHarlingen} by directly solving the
real-axis Eliashberg equations (EEs) in the half-filling case.

Incidentally, this procedure is much more complicated than the
usual one, which consists in solving the EEs in the imaginary-axis
formulation and then continuing the solution to the real axis.
Nevertheless, it is much more general and can be used at every
temperature, whilst the second approach is meaningful only at very
low temperatures (in principle, $T \rightarrow 0$). Moreover, we
will show that, at a finite but still very low temperature
($T$~=~2~K) the imaginary-axis procedure gives results whose
agreement with those directly obtained from the real-axis EEs is
not equally good for all symmetries.

We will also present the theoretical conductance curves obtained
at various temperatures, and with increasing values of the
coupling constant. It will be shown that large values of the
isotropic coupling constant produce some characteristic and
well-recognizable features of the normalized conductance curves,
which are actually observed experimentally.

Moreover, we will investigate the effect of different amounts of
{\em non-magnetic} impurities, in both the unitary and non-unitary
limits, on the tunneling normalized conductance curves, and
discuss the results for all the symmetries of the order parameter.

Finally, we will try a comparison of the theoretical tunneling
curves with experimental data appeared in literature and
concerning some high-$T\ped{c}$ compounds.

\section{Eliashberg equations for different pair symmetries}

In this section we calculate the theoretical normalized
conductance for a tunnel junction within the framework of
Migdal-Eliashberg theory. To do so, we solve the generalized
real-axis EEs \cite
{Eliashberg,Carbotte,AllenMitr} for the renormalization function $Z(\omega ,%
{\bf k})$ and the order parameter $\Delta (\omega ,{\bf k})$. As
well known, in the real-axis formalism the EEs take the form of a
set of coupled integral
equations, whose kernels contain the retarded electron-boson interaction $%
\alpha ^{2}F(\Omega ,{\bf k},{\bf k}^{\prime })$, the Coulomb
pseudopotential $\mu ^{\ast }\left( {\bf k},{\bf k}^{\prime
}\right) $, and the energy of the carriers $\varepsilon _{{\bf
k}}$, measured with respect to the bare band chemical potential
\cite{Rieck,Strinati,JiangCarb}. In the following,  we will use a
single-band approximation, and we will restrict our discussion to
the 2-dimensional case, thus referring, for example, to the $ab$
planes of layered superconductors and neglecting the band
dispersion and the gap in the $c$ direction. For simplicity, we
will assume that the Fermi surface is a circle in the $ab$ plane
\cite{Rieck} and that the wavevectors ${\bf k}$ and ${\bf
k}^{\prime }$ are completely determined by the respective
azimuthal angles $\phi $ and $\phi ^{\prime }$, since their length
is, as usual, taken equal to k$_{{\rm F}}$.

To allow for different symmetries of the pair state, we expand the
superconducting spectral function $\alpha ^{2}F(\Omega ,\phi ,\phi
^{\prime })$ and the Coulomb pseudopotential $\mu ^{\ast }(\phi
,\phi
^{\prime })$ in terms of basis functions $\psi _{{\rm is}}$ and $\psi _{{\rm %
an}}$, where the subscripts mean  {\it isotropic} and {\it
anisotropic}, respectively. At the lowest order we suppose $\alpha
^{2}F(\Omega ,\phi ,\phi ^{\prime })$ and $\mu ^{\ast }(\phi
,\phi ^{\prime })$ to contain {\it separated} isotropic and
anisotropic contributions, as in the following expressions:
\begin{equation}
\alpha ^{2}F(\Omega ,\phi ,\phi ^{\prime })=%
\alpha^{2}F (\Omega )_{{\rm is}}\psi _{{\rm is}}\left( \phi \right) \psi%
_{{\rm is}}\left(\phi ^{\prime }\right) +\alpha^{2}F (\Omega )_{{\rm an}}\psi _{{\rm an}}%
\left( \phi \right) \psi _{{\rm an}}\left( \phi ^{\prime }\right)
\label{alfa^2}
\end{equation}
\vspace{-9mm}
\begin{equation}
\mu ^{\ast }(\phi ,\phi ^{\prime })=\mu _{{\rm is}}^{\ast }\psi _{{\rm is}%
}\left( \phi \right) \psi _{{\rm is}}\left( \phi ^{\prime }\right) +\mu _{%
{\rm an}}^{\ast }\psi _{{\rm an}}\left( \phi \right) \psi _{{\rm
an}}\left( \phi ^{\prime }\right) .  \label{mu*}
\end{equation}

The basis function $\psi _{{\rm is}}\left( \phi \right) $ and $\psi _{{\rm an%
}}\left( \phi \right) $ are chosen as follows:
\begin{eqnarray}
\psi _{{\rm is}}\left( \phi \right) &=&1  \nonumber \\
\psi _{{\rm an}}\left( \phi \right) &=&\left\{
\begin{array}{l}
\sqrt{2}\cos \left( 2\phi \right) \\ 8\sqrt{2/35}\cos ^{4}\left(
2\phi \right) \\ 2\sqrt{2/3}\cos ^{2}\left( 2\phi \right)
\end{array}
\right.\hspace{10mm}
\begin{array}{l}
(d-wave) \\ (anisotropic~\emph{s}) \\
(extended~\emph{s})
\end{array}
\label{psi}
\end{eqnarray}

We search for solutions of the EEs having the form:
\begin{eqnarray}
\Delta (\omega ,\phi ) &=&\Delta _{{\rm is}}(\omega )\pm\Delta _{{\rm an}%
}(\omega )\psi _{{\rm an}}\left( \phi \right)  \label{DeltaZeta}
\\
Z(\omega ,\phi ) &=&Z_{{\rm is}}(\omega )\pm Z_{{\rm an}}(\omega )\psi _{{\rm an%
}}\left( \phi \right) .  \nonumber
\end{eqnarray}

We assume as an ``ansatz'' that the negative sign in the above
expressions is used \emph{only} for the \emph{extended s}-wave
symmetry, while in all other cases the positive sign is chosen. In
doing so one recovers in the BCS limit the form of $\Delta$ given
in Ref.\cite{VanHarlingen}. Notice that the choice of the sign in
the expression for $Z(\omega, \phi)$ has no effect. In fact, using
eqs.(\ref{DeltaZeta}) makes the Eliashberg equations for $\Delta
(\omega ,\phi )$ and $
Z(\omega ,\phi )$ split into four equations for $\Delta _{{\rm is}}(\omega )$%
, $Z_{{\rm is}}(\omega )$, $\Delta _{{\rm an}}(\omega )$ and $Z_{{\rm an}%
}(\omega )$. The equation for $Z_{{\rm an}}(\omega )$ is a homogeneous
integral equation whose only solution in the weak-coupling regime is $Z_{%
{\rm an}}(\omega )\equiv 0$. Even though in the strong-coupling limit a
non-zero solution could exist above a certain coupling strength threshold,
we do not consider here this rather exotic case and then we assume that the
stable solution corresponds to $Z_{{\rm an}}(\omega )\equiv 0$ for all
couplings \cite{Musaelian,JohnRenato}. Consequently, we will not write the
equation for $Z_{{\rm an}}(\omega )$.

In writing the remaining equations for $\Delta _{{\rm is}}(\omega )$, $Z_{%
{\rm is}}(\omega )$, $\Delta _{{\rm an}}(\omega )$, we insert an
additional term which takes into account the presence of
non-magnetic impurities \cite{JiangCarbDynes}. Actually, in the
following numerical solutions we will consider this term only
where explicitly specified, and we will disregard it otherwise.
The EEs are then written in the form:

\vspace{-3mm}
\begin{eqnarray}
\Delta _{{\rm is}}(\omega )Z_{{\rm is}}(\omega ) &=&\frac{1}{2\pi }%
\int_{0}^{2\pi }{\rm d}\phi \int_{0}^{+\infty }\mathrm{Re}\left[ P_{{\rm is}}(\omega ^{\prime },\phi )\right] \left[ K_{%
{\rm is+}}(\omega ,\omega ^{\prime },\phi )-\mu _{{\rm is}}^{\ast
}\theta \left( \omega _{c}-\omega ^{\prime }\right) \right]{\rm
d}\omega^{\prime }   \nonumber \\
&& +{\rm i}\pi \Gamma \frac{\overline{P}_{{\rm is}}(\omega )}{C^{2}+\overline{P}_{{\rm %
an}}^{2}(\omega )+\overline{P}_{{\rm is}}^{2}(\omega )+\overline{N}_{{\rm is}%
}^{2}(\omega )} \label{EE1}
\end{eqnarray}
\vspace{-3mm}
\begin{eqnarray}
\Delta _{{\rm an}}(\omega )Z_{{\rm is}}(\omega ) &=&\frac{1}{2\pi }%
\int_{0}^{2\pi }{\rm d}\phi \int_{0}^{+\infty }\mathrm{Re} \left[ P_{{\rm an}}(\omega ^{\prime },\phi )\right] \left[ K_{%
{\rm an+}}(\omega ,\omega ^{\prime },\phi )-\mu _{{\rm an}}^{\ast
}\theta \left( \omega _{c}-\omega ^{\prime }\right) \psi _{{\rm
an}}\left( \phi \right) \right]{\rm d}\omega
^{\prime }  \nonumber \\
&&+ {\rm i}\pi \Gamma \frac{\overline{P}_{{\rm an}}(\omega )}{C^{2}+\overline{P}_{{\rm %
an}}^{2}(\omega )+\overline{P}_{{\rm is}}^{2}(\omega )+\overline{N}_{{\rm is}%
}^{2}(\omega )}  \label{EE2}
\end{eqnarray}
\vspace{-3mm}
\begin{eqnarray}
\left[ 1-Z_{{\rm is}}(\omega )\right] \omega &=&\frac{1}{2\pi }%
\int_{0}^{2\pi }{\rm d}\phi  \int_{0}^{+\infty }\mathrm{Re} \left[ N_{{\rm is}}(\omega ^{\prime },\phi )\right] K_{{\rm is-%
}}(\omega ,\omega ^{\prime },\phi ){\rm d}\omega
^{\prime }  \label{EE3} \\
&&- {\rm i}\pi \Gamma \frac{\overline{N}_{{\rm is}}(\omega )}{C^{2}+\overline{P}_{{\rm %
an}}^{2}(\omega )+\overline{P}_{{\rm is}}^{2}(\omega )+\overline{N}_{{\rm is}%
}^{2}(\omega )} .  \nonumber
\end{eqnarray}

\noindent{where $\omega _{{\rm c}}$, which appears in
eqs.(\ref{EE1}, \ref{EE2}), is a cutoff energy, and the functions
$P\ped{is}(\omega, \phi)$, $P\ped{an}(\omega, \phi)$ and
$N\ped{is}(\omega, \phi)$ are defined as follows: }

\vspace{-3mm}
\begin{equation}
P_{{\rm is}}(\omega ,\phi )=\frac{\Delta _{{\rm is}}(\omega
)}{\sqrt{\omega
^{2}-\left[ \Delta _{{\rm is}}(\omega )+\Delta _{{\rm an}}(\omega )\psi _{%
{\rm an}}\left( \phi \right) \right] ^{2}}}  \label{Pis}
\end{equation}

\vspace{-3mm}
\begin{equation}
P_{{\rm an}}(\omega ,\phi )=\frac{\Delta _{{\rm an}}(\omega )\psi _{{\rm an}%
}\left( \phi \right) }{\sqrt{\omega ^{2}-\left[ \Delta _{{\rm
is}}(\omega
)+\Delta _{{\rm an}}(\omega )\psi _{{\rm an}}\left( \phi \right) \right] ^{2}%
}}  \label{Pan}
\end{equation}

\vspace{-3mm}
\begin{equation}
N_{{\rm is}}(\omega , \phi )= \frac{\omega }{%
\sqrt{\omega ^{2}-\left[ \Delta _{{\rm is}}(\omega )+\Delta _{{\rm an}%
}(\omega )\psi _{{\rm an}}\left( \phi \right) \right] ^{2}}}
\label{Nis}
\end{equation}

\noindent{while} \vspace{-3mm}
\begin{equation}
\overline{P}\ped{is, an}(\omega)=\frac{1}{2%
\pi}\int_{0}^{2 \pi}P\ped{is, an}(\omega, \phi)\:{\rm d
}\phi\label{Paverage}
\end{equation}
\noindent{and} \vspace{-3mm}
\begin{equation}
\overline{N}\ped{is}(\omega)=\frac{1}{2 \pi}\int_{0}^{2%
\pi}N\ped{is}(\omega, \phi)\:{\rm d }\phi \label{Naverage}.
\end{equation}

The functions $K_{{\rm is,an\pm }}(\omega ,\omega ^{\prime },\phi
)$ in the integral over $\omega ^{\prime }$ are given by
\begin{eqnarray}
K_{{\rm is,an\pm }}(\omega ,\omega ^{\prime },\phi ) &=&\int_{0}^{+\infty }%
{\rm d}\Omega \: \alpha ^{2}F(\Omega )_{{\rm is,an}}\:\psi _{{\rm is,an}%
}\left( \phi \right) \cdot  \label{Kappa} \\
&&\left[ \frac{f\left( -\omega ^{\prime }\right) +n\left( \Omega \right) }{%
\Omega +\omega ^{\prime }+\omega +{\rm i}\delta }\pm \frac{f\left(
-\omega ^{\prime }\right) +n\left( \Omega \right) }{\Omega +\omega
^{\prime }-\omega -{\rm i}\delta }\mp \frac{f\left( \omega
^{\prime }\right) +n\left( \Omega \right) }{\Omega -\omega
^{\prime }+\omega +{\rm i}\delta }\mp \frac{f\left( \omega
^{\prime }\right) +n\left( \Omega \right) }{\Omega -\omega
^{\prime }-\omega -{\rm i}\delta }\right]  \nonumber
\end{eqnarray}

\noindent{where $f\left( \omega \right) $ and $n\left( \Omega%
\right) $ are the Fermi and Bose distributions, respectively.}

As anticipated above, the last term on the right-hand side of
equations (\ref{EE1}-\ref{EE3}) allows for the contribution of
scattering from impurities. It contains the parameters $\Gamma $
and $C$ which are defined through the relations $\Gamma =n_{{\rm
I}}/N(0)\pi ^{2} $ and $C$=$\cot(\delta _{0})$, where $n_{{\rm
I}}$ is the impurity concentration, $N(0)$ is the value of the
normal DOS at the Fermi energy and $\delta _{0}$ is the
scattering phase shift \cite{JiangCarbDynes}.

Notice that, in the case we want to obtain a $s+{\rm i}d$
solution, we
simply need to replace the denominator of $P_{{\rm an}}(\omega ,\phi )$, $P_{%
{\rm is}}(\omega ,\phi )$ and $N\ped{is}(\omega ,\phi )$ with
$\sqrt{ \omega ^{2}-\left[ \Delta _{{\rm s}}^{2}(\omega )+\Delta
_{{\rm d}}^{2}(\omega )\psi ^{2}_{{\rm d}}\left( \phi \right)
\right] }$, where the generic subscripts `is' and `an' have been
replaced by `s' and `d', respectively.

To simplify the problem of numerically solving these equations,
we put $\alpha ^{2}F(\Omega )_{{\rm an}}=g\cdot \alpha%
^{2}F(\Omega )_{{\rm is}}$ where $g$ is a constant
\cite{JohnRenato,Dolgov}. As a consequence, the electron-boson
coupling constants for the {\it isotropic}-wave channel and the
{\it anisotropic}-wave one, which are given by $\lambda _{{\rm
is}}$=$2\int_{0}^{+\infty }{\rm d}\Omega \: \alpha^{2}F(\Omega
)_{{\rm is}}/\Omega $ and $\lambda _{{\rm an}}$=$\left( 1/\pi
\right) \int_{0}^{2\pi }{\rm d}\phi \: \psi _{{\rm an}}^{2}\left(
\phi \right) \int_{0}^{+\infty }{\rm d}\Omega \:
\alpha^{2}F(\Omega )_{{\rm an}}/\Omega$ respectively, result to
be proportional: $\lambda _{{\rm an}}=g\cdot \lambda _{{\rm is}}$.

The real-axis generalized Eliashberg equations
(\ref{EE1}-\ref{EE3}) are numerically solved by using an iterative
procedure, which stops when the real and imaginary values of the
functions $\Delta _{{\rm is}}(\omega ),$ $\Delta _{{\rm
an}}(\omega )$ and $Z_{{\rm is}}(\omega )$ at a new iteration
differ less than $1\cdot 10^{-3}$ with respect to the previous
ones. This usually happens after a number of iterations between
10 and 15.

By starting from the real-axis solutions $\Delta _{{\rm is}}(\omega )$ and $%
\Delta _{{\rm an}}(\omega )$, we then calculate the normalized
conductance $G\ped{n}(V)$ of a tunnel junction, which is given by
the convolution of the quasiparticle density of states $N(\omega)=\mathrm{Re}%
\left[\mathrm{sgn}(\omega) \overline{N}\ped{is}(\omega)\right] $
with the Fermi distribution function: 
\begin{equation}\vspace{-2mm}
G\ped{n}(V)=\frac{1}{e} \int_{-\infty }^{+\infty }{\rm d}\omega \:
N(\omega+eV)\cdot \left[f(\omega)-f(\omega+eV )\right].
\label{N(omega)}\vspace{1mm}
\end{equation}

Of course, this is the simplest possible choice. We are aware
that, when anisotropic superconductors are involved, the
normalized conductance should have a much more complex form to
take into account its dependence on the junction's geometry and on
some other factors such as, for example, the direction of the
incident current and the angle $\alpha$ between the normal to the
interface and the crystallographic $a$ axis \cite{Tanaka}. Here we
restrict ourselves to the simplest case of plane interface and
normal current and we also put $\alpha=0$.

The main features of the resulting $G\ped{n}(V)$ curves will be
discussed in the following sections.

As already pointed out in the introduction, the direct numerical
solution of the real-axis EEs described so far is a quite
complicated and time-consuming procedure. However, as well known,
the EEs admit an imaginary-axis formulation, in which integrals
are replaced by summations
over the integer index of the so-called Matsubara frequencies $\omega _{%
{\rm n}}$. The numerical solution of the EEs in this formalism is
therefore a
much simpler task, which gives the real discrete functions $\Delta \left( {\rm i}%
\omega _{{\rm n}}\right) $ and $Z({\rm i}\omega _{{\rm n}})$, that
can be analytically continued to the real axis by means of
Pad\'{e} approximants \cite{VidbergSerene}. In principle, this
analytical continuation is meaningful only at very low
temperatures. Therefore, one could wonder if, and in what
temperature range, this method is able to give results in
accordance with those obtained by starting from the real-axis EEs.
A comparison  between the density of states (DOS) obtained by the
two techniques was made for the $s$-wave symmetry and for small
values of the coupling constant
($\lambda<2$)~\cite{VidbergSerene}, and a substantial agreement
was found. Here we compare the normalized tunneling conductance
curves calculated at $T=2$~K not only in the $s$-wave symmetry
for greater values of $\lambda$, but also in the other symmetries
analyzed here. The results will be discussed in the next section.

\begin{figure}[t]\vspace{-3mm}
\begin{minipage}{0.35\textwidth}
\includegraphics[keepaspectratio,width=5.2cm]{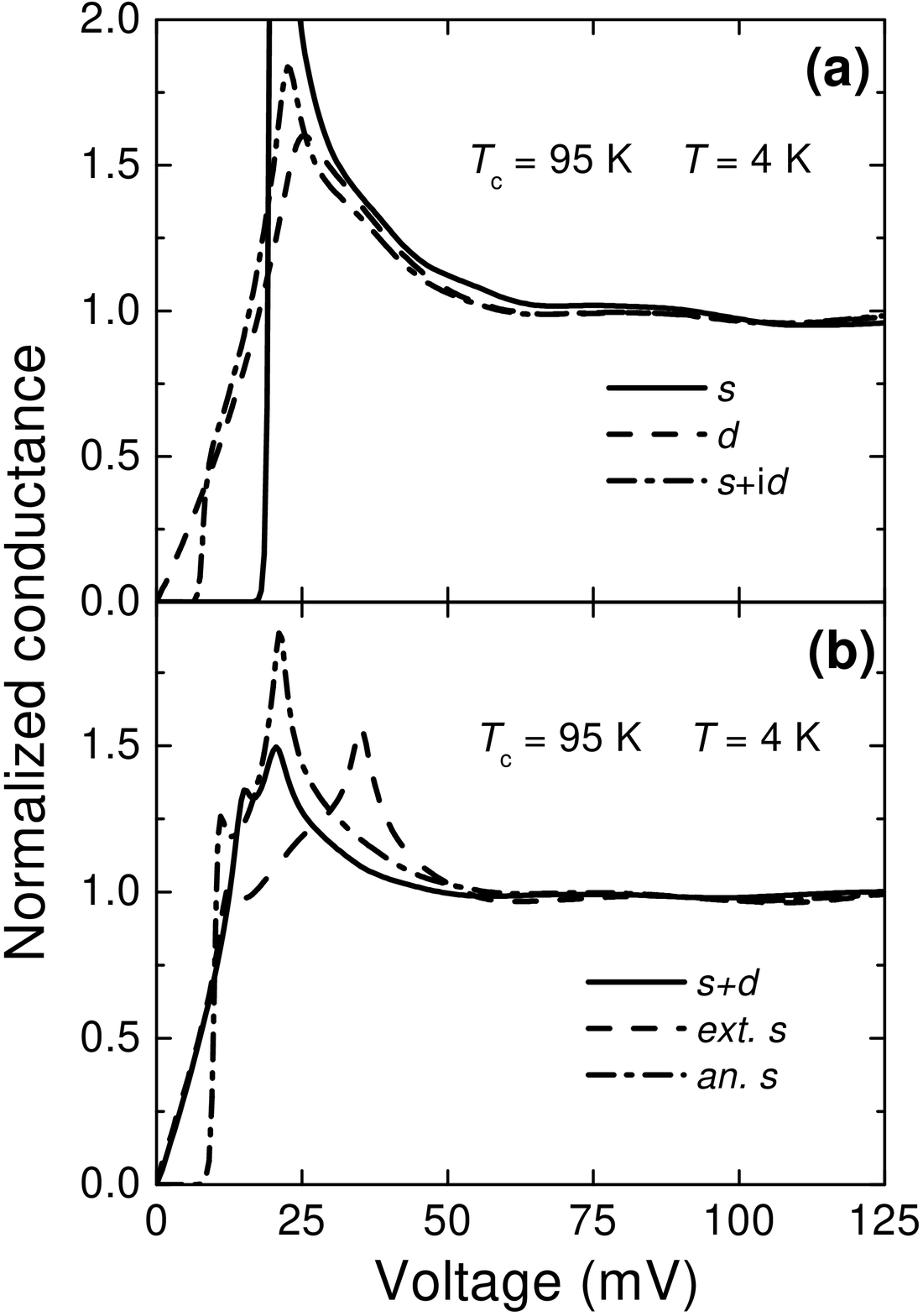}
\caption{{\small Theoretical normalized conductance curves for
$\mu ^{*}\:$=~0 at $T\:$=~4~K in various symmetries of the order
parameter: {\it s}, {\it d}, {\it s}+i{\it d} (a) and {\it
s}+{\it d}, {\it \ extended s}, {\it anisotropic s} (b).}}
\end{minipage}
\hfill
\begin{minipage}{0.6\textwidth}
\vspace{-5mm}
\includegraphics[keepaspectratio,width=1.02\textwidth]{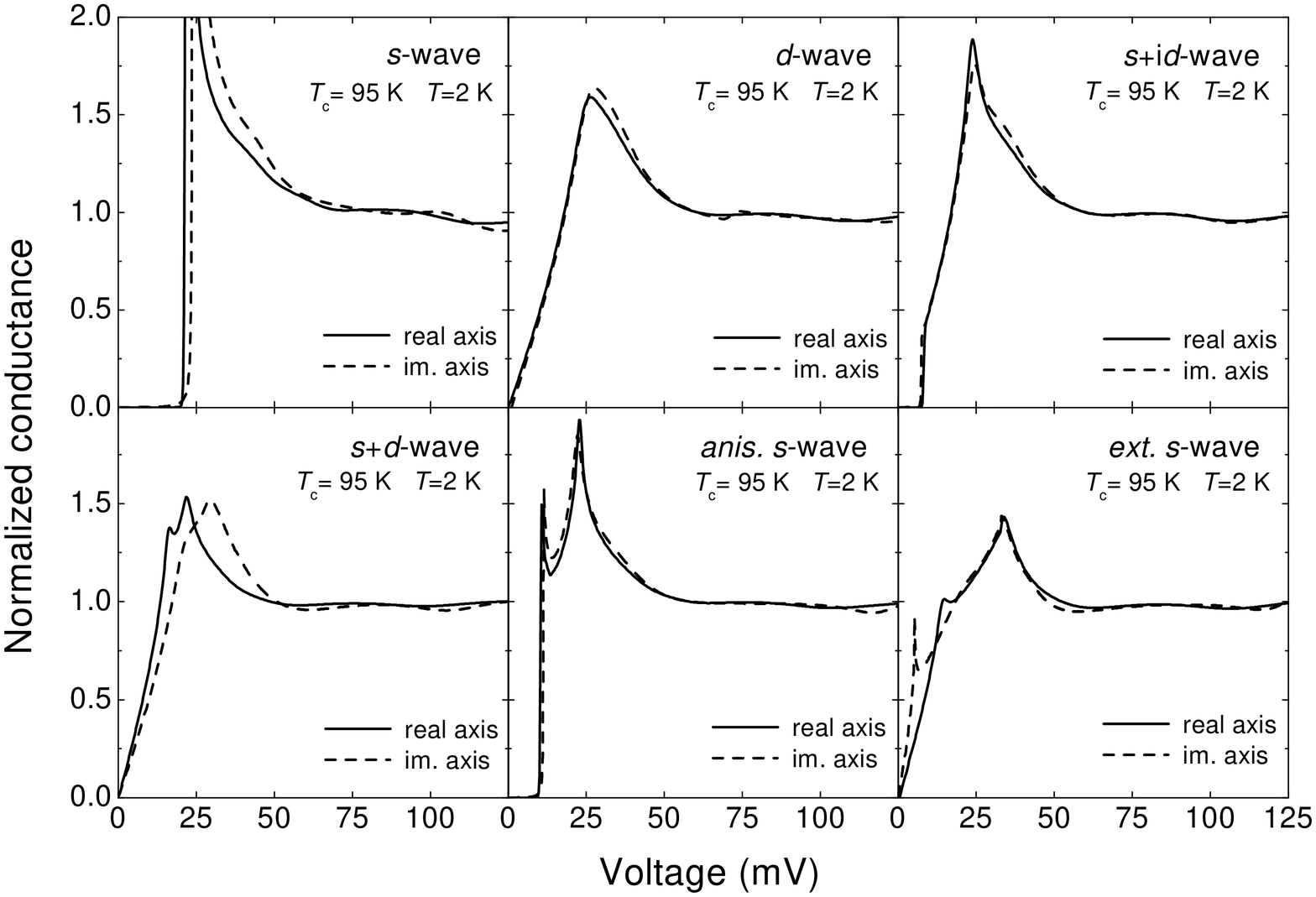}
\vspace{-6mm}\caption{{\small Comparison of the normalized
tunneling conductance curves obtained from the real-axis solution
of the Eliashberg equations, with those obtained by the analytical
continuation of the imaginary-axis solutions, at $T$=2~K, in all
the symmetries considered.}}
\end{minipage}
\end{figure}

\section{Theoretical Results}
The numerical solution of the EEs (\ref{EE1})-(\ref{EE3}) requires
the knowledge of the function $\alpha^{2}F(\Omega)\ped{is}$ which
appears in the expressions of $K\ped{is,an\pm}$ and
$\lambda\ped{is, an}$. Even though we are not referring to a
particular material, we can make a choice for
$\alpha^{2}F(\Omega)\ped{is}$ by taking advantage of the fact that
its detailed shape has little influence on the final solution,
which instead depends, as observed
elsewhere~\cite{Varelogiannis2}, on
the quantity $\omega _{\log }=$ $\exp \left( \frac{2}{\lambda%
}\int_{0}^{+\infty }d\Omega \frac{\alpha ^{2}F\left( \Omega%
\right) }{\Omega }\ln \Omega \right) $. Since in this section we
are only interested in calculating theoretical curves, we can
take: 
\begin{equation}
\alpha ^{2}F(\Omega )_{{\rm is}}=\frac{\lambda _{{\rm
is}}}{\lambda }\alpha ^{2}F(\Omega )_{{\rm Bi2212}}.
\label{alfaBi}
\end{equation}

Here $\alpha ^{2}F(\Omega )_{{\rm Bi2212}}$ is the electron-boson
spectral function determined in a previous paper by inversion of
the $s$-wave EEs starting from tunneling data obtained on
Bi$_2$Sr$_2$CaCu$_2$O$_{8+\delta}$ (Bi-2212) break junctions
\cite{PhysC97}, and $\lambda$ is the corresponding coupling
constant. The values of $\lambda\ped{is}$ and that of
$\lambda\ped{an}$ are chosen to have
$T\ped{c}\equiv\max(T\ped{c}^{\mathrm{is}},T\ped{c}^{\mathrm{an}})$~=~95~K
(which, in most of our calculations, is chosen as a representative
value for high-$T\ped{c}$ superconductors).

From now on, we will put the Coulomb pseudopotential $\mu ^{\ast
}$ to zero in all the equations of the previous section. It will
be shown in the next section that this does not limit the
generality of the results.

As already pointed out, we search for a solution of the real-axis
Eliashberg equations which contains an isotropic and an
anisotropic part, as shown in equation (\ref{DeltaZeta}).
Actually, the choice of the coupling constants $\lambda _{{\rm
is}}$ and $\lambda _{ {\rm an}}$ to be used for the numerical
solution  can affect the symmetry of the order parameter $\Delta
(\omega ,\phi )$, which can show either pure {\it isotropic}-wave
or {\it anisotropic}-wave symmetry, or a mixed symmetry. Moreover,
for some particular values of the coupling constants, the final
symmetry of the order parameter depends on the starting values of
$\Delta _{{\rm is}}(\omega )$ and $\Delta _{{\rm an }}(\omega )$.

Results with the expected symmetry, and compatible with the value
of the critical temperature we have chosen above ($T_{{\rm c}}$=
95~K) are obtained with the values of the coupling constants
reported in Table I, where also the corresponding isotropic and
anisotropic critical temperatures are shown.
\begin{table}[b]
\begin{center}
\def \D {\rule[-1ex]{0pt}{3.5ex}}
\begin{tabular}{|c|c|c|c|c|}
\hline \D symmetry & $\lambda\ped{is}$ & $\lambda\ped{an}$ %
& $T\ped{c}^{\mathrm{is}}$~(K) & $T\ped{c}^{\mathrm{an}}$~(K) \\
\hline \hline \D\emph{s} & 3.15 & - & 95 & - \\ \hline\D\emph{d} &
2 & 2.3 & - & 95 \\ \hline\D\emph{s}+i\emph{d} & 2 & 2.3 & 66 & 95
\\ \hline\D\emph{s}+\emph{d} & 1.25 & 1.74 & 40 & 95 \\ \hline
\D\emph{anis. s} & 2 & 2.29 & 63 & 95 \\ \hline\D\emph{ext. s} & 1
& 1.54 & 32 & 95 \\ \hline
\end{tabular}
\end{center}
\caption{Values of the coupling constants used in the calculations
of Section~III and corresponding critical temperatures for all the
gap symmetries considered. }
\end{table}

It is worthwhile to remark that, in all cases apart from the
$s$-wave one, different couples of $\lambda _{{\rm is}}$ and
$\lambda _{{\rm an}}$ values can give rise to the same critical
temperature and the same symmetry and, therefore, Table~I only
presents one of the possible choices. These coupling constants
will be used in all the calculations presented in this section.

Figure~1 reports the theoretical normalized tunneling conductance
curves $G\ped{n}(V)$ at $T=4$~K in the six different symmetries
analyzed (the corresponding values of the coupling constants are
those reported above). In all the symmetries apart from the
$s$-wave one, a conductance excess below the gap is obtained.
However, in all cases the normalized conductance is zero at zero
voltage. As we will discuss later, the zero-bias anomaly which is
often experimentally observed could in fact be reproduced, in the
framework of the Eliashberg theory, by taking into account the
scattering from impurities, which has been disregarded here.

\begin{figure}[t]\hspace{-2mm}
\begin{minipage}{0.3\textwidth}
\vspace{-9mm}
\includegraphics[keepaspectratio,width=\textwidth]{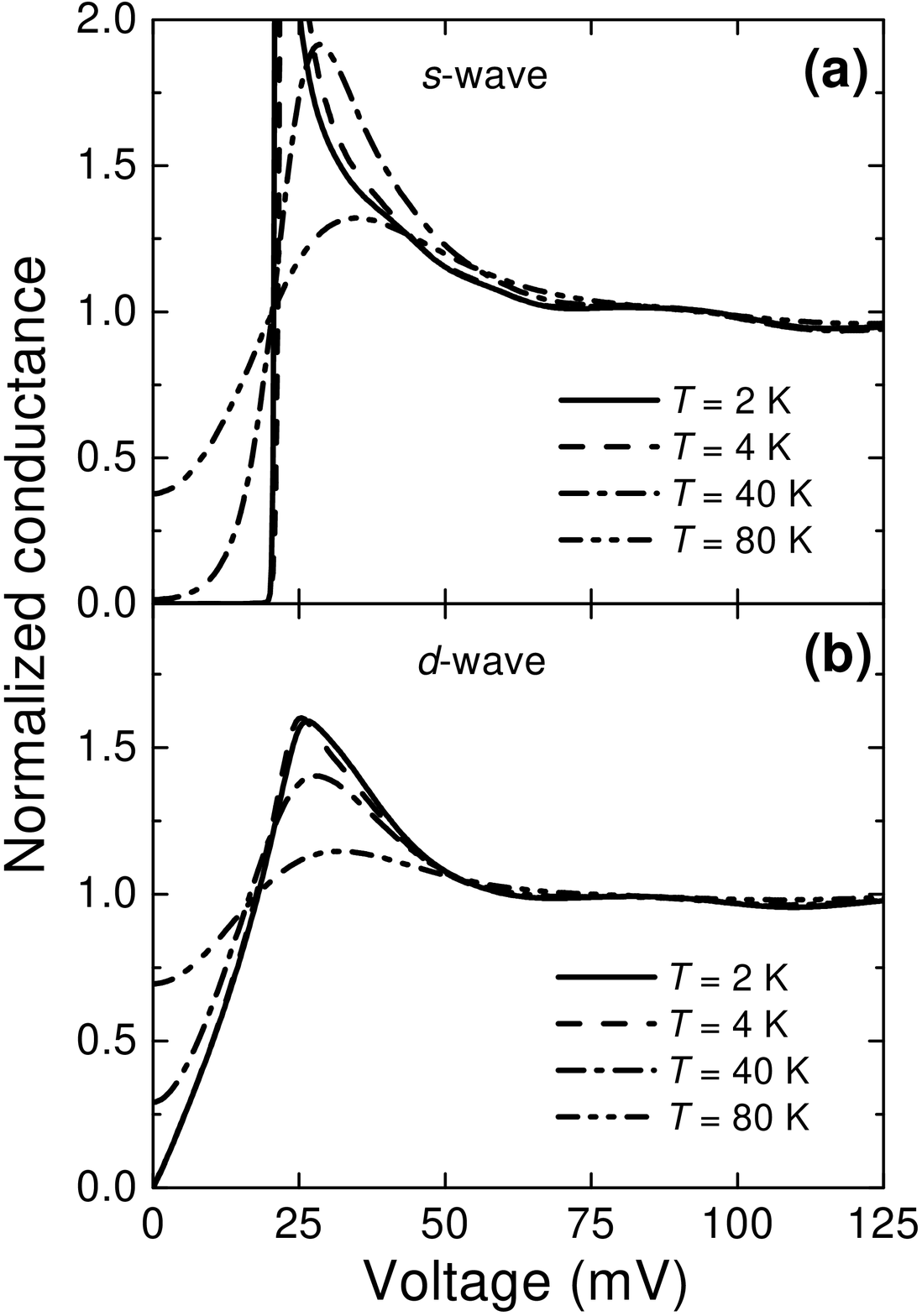}
\caption{{\small Theoretical normalized conductance curves for
$\mu ^{\ast }\:$=~0 at $T$\@=2, 4, 40 and 80~K in the $s$-wave
symmetry (a) and in the $d$-wave symmetry (b).}}
\end{minipage}
\hfill
\begin{minipage}{0.3\textwidth}
\vspace{-5mm}
\includegraphics[keepaspectratio,width=\textwidth]{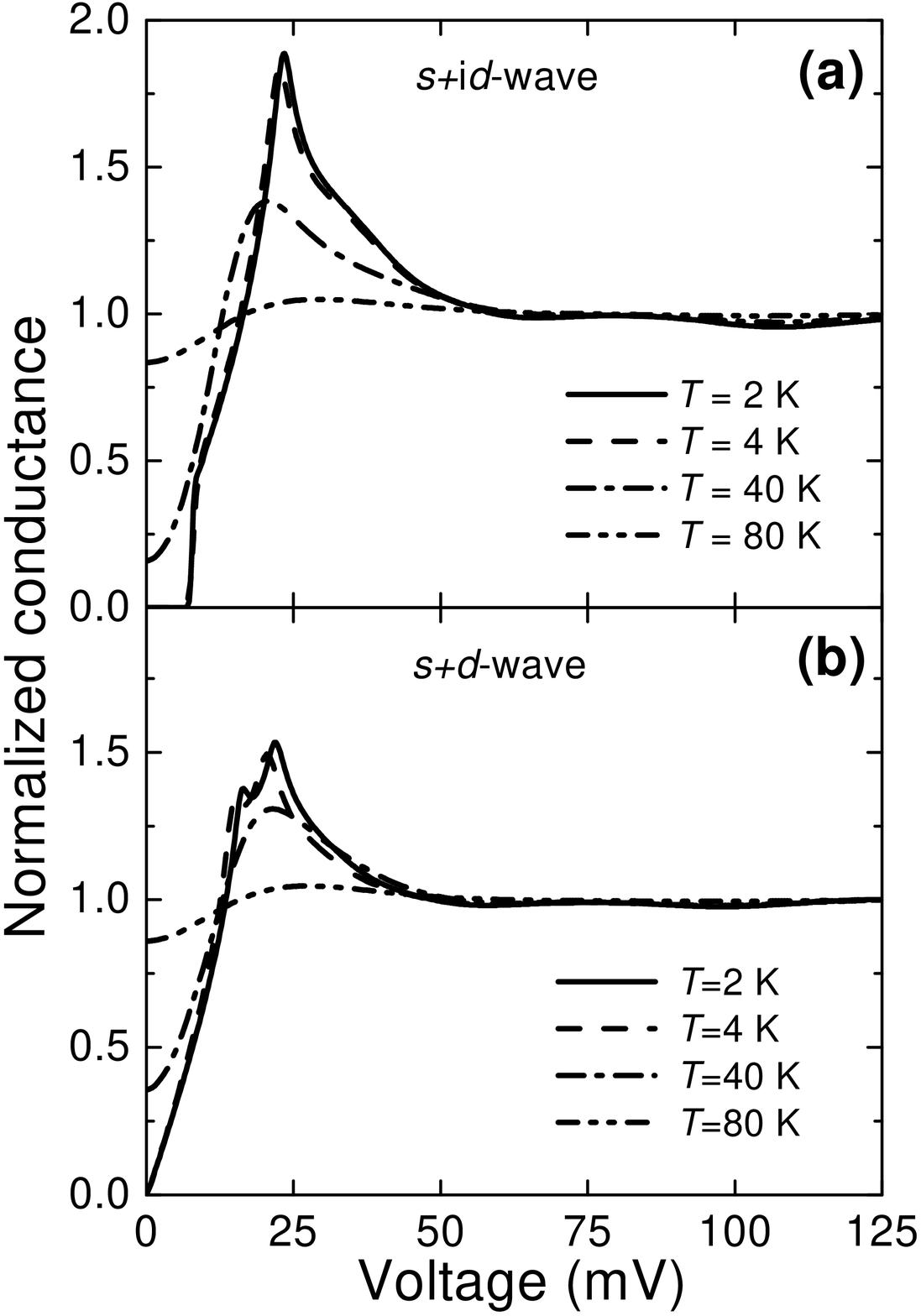}
\caption{{\small Theoretical normalized conductance curves for
$\mu ^{\ast }\:$=~0 at $T$\@=2, 4, 40 and 80~K in the
$s$+i$d$-wave symmetry (a) and in the $s$+$d$-wave symmetry (b).}}
\end{minipage}
\hfill
\begin{minipage}{0.3\textwidth}
\vspace{-5mm} \hspace{-3mm}
\includegraphics[keepaspectratio,width=\textwidth]{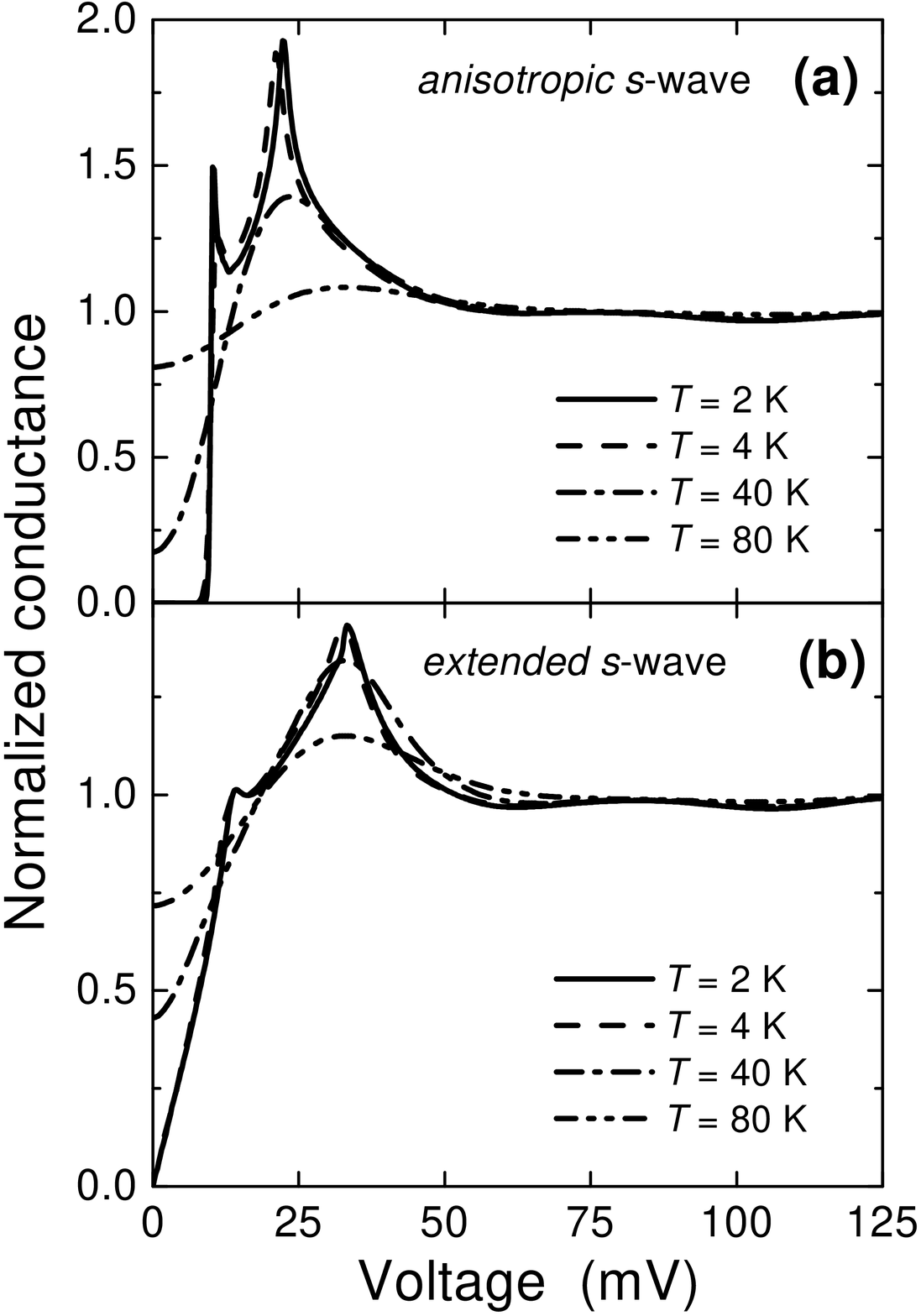}
\caption{{\small Theoretical normalized conductance curves for
$\mu ^{\ast }\:$=~0 at $T$\@=2, 4, 40 and 80~K in the {\it
anisotropic} $s$-wave symmetry (a) and in the {\it extended}
$s$-wave symmetry (b).}}
\end{minipage}
\end{figure}

As anticipated in the previous section, our choice of directly
solving the real-axis EEs can be motivated by the fact that the
analytical continuation of the solutions $\Delta (%
{\rm i}\omega _{n})$ and $Z({\rm i}\omega\ped{n})$ of the EEs in
the imaginary-axis formulation is correct only at very low
temperature. The meaning of this `very low' can be clarified by
comparing the results of the two approaches at various
temperatures. The result of such a comparison is that, even at the
liquid helium \phantom{temperature}

\begin{figure}[t]
\begin{minipage}{0.3\textwidth}
\vspace{-5mm}
\includegraphics[keepaspectratio,width=\textwidth]{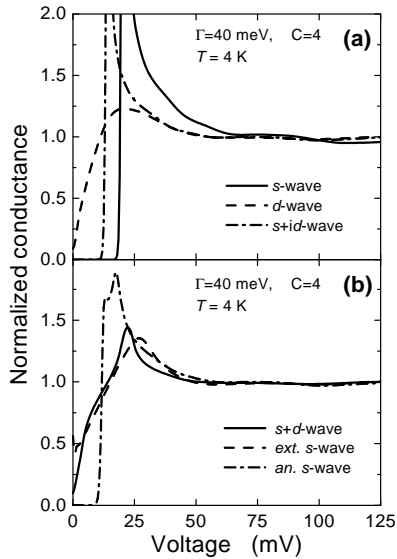}
\vspace{-8mm} \caption{{\small Theoretical normalized conductances
curves at $T\:$=4~K in various gap symmetries, for $\mu ^{\ast
}\:$=~0 and in the presence of scattering from impurities in the
non-unitary limit ($\Gamma$~=~40~meV and $C$~=~4).}}
\end{minipage}
\hfill
\begin{minipage}{0.65\textwidth}
\vspace{3mm}temperature (at which most of the experimental data
are obtained), the analytical continuation gives rise to some
sensible deviations from the real-axis solution. As shown in
Fig.~2, at a lower temperature ($T=2$~K), the agreement between
the theoretical $G\ped{n}(V)$ curves obtained by the two
procedures is good for the $d$-wave, $s+\mathrm{i}d$-wave and
\emph{anisotropic s}-wave symmetries. In the {\it extended }$s$
case, the `imaginary-axis' curve agrees very well with the
`real-axis' one, except that in the low-voltage regime. In the
($s+d$)-wave symmetry, the peak of the imaginary-axis solution is
broader than that of the real-axis one, and displaced of about
+8~mV. Finally, in the $s$-wave case, the analytical continuation
does not work well, since a shift of +3~mV of the conductance
curve is produced. Further checks demonstrate that the agreement
gets worse at the increase of $\lambda$, while for sufficiently
small values of the coupling constant ($\lambda <2$) the two
curves coincide, as found in Ref.~\cite{VidbergSerene}.

\hspace{3mm} Let us now discuss what happens to the conductance
curves when the temperature is increased. Figures~3, 4 and~5 show
the normalized tunneling conductance curves calculated for all the
symmetries at $T$~=~2, 4, 40, 80~K. In general, as expected,
increasing the temperature results in a smoothing and broadening
of the peaks. The smoothing is particularly evident in the case of
the {\em anisotropic }$s$ symmetry, in which the low-voltage peak,
which is well defined at $T$~=~2~K, is much less evident at 4~K
(its height is reduced from 1.5 to 1.25). At the increase of the
temperature the curves corresponding to mixed symmetries become
more and more similar to one another and to the $d$-wave curve.
This is due both to the thermal broadening and to the fact that,
with our choices of $\lambda\ped{is}$ and $\lambda\ped{an}$,
$T\ped{c}^{\mathrm{is}}$ is always less than
$T\ped{c}^{\mathrm{an}}$ and therefore the isotropic gap component
disappears before the anisotropic one. The $s$-wave curve remains
instead clearly distinct.
\end{minipage}\vspace{-11mm}
\end{figure}

The theoretical curves presented and discussed so far were
obtained without inserting in the EEs the additional term taking
into account the scattering from non-magnetic impurities. When
this effect is allowed for in solving the real-axis EEs, one finds
that in the general non-unitary case ($C \neq 0$) it mostly
affects the $d$-wave component of the order parameter. By
comparing the curves shown in Fig.~6 to those of Fig.~1, we can
see that the peak of the $d$-wave tunneling conductance curve is
lowered and broadened, while the $s+{\mathrm i}d$ curve becomes
practically indistinguishable from an $s$-wave one. In the $s+d$
and \emph{extended}~$s$ cases the low-energy peak disappears and a
zero-bias is produced, which is greater in the latter case.
Finally, in the {\it anisotropic}~$s$ case the two peaks are
closer than in the absence of impurities and the gap is shifted to
the left, but the general shape above and below the gap is
conserved.

\begin{figure}[t]\hspace{-2mm}
\begin{minipage}{0.3\textwidth}
\vspace{-1mm}
\includegraphics[keepaspectratio,width=\textwidth]{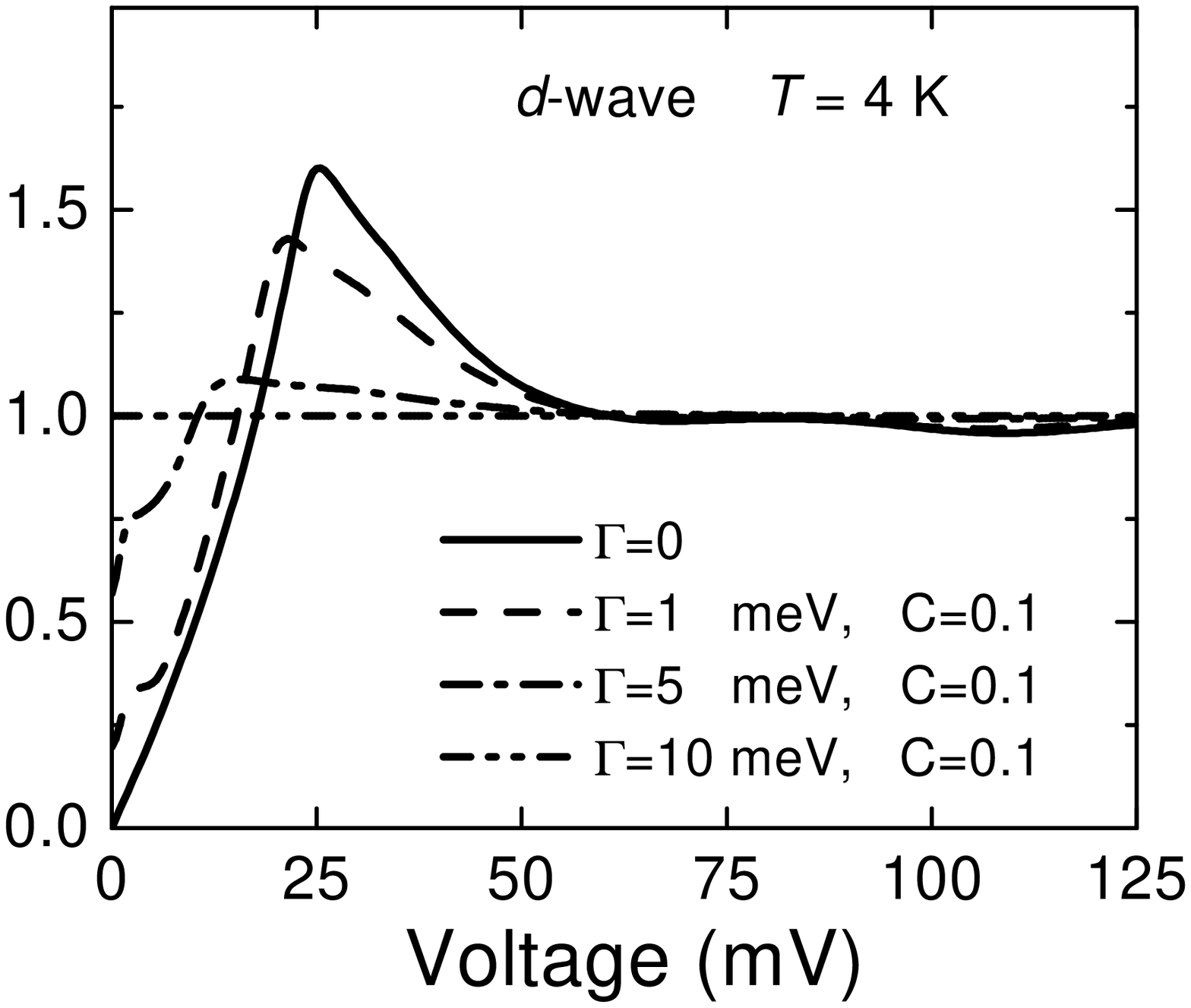}
\caption{{\small Theoretical normalized conductance curves for
$\mu ^{*}\:$=~0 in the $d$-wave symmetries at $T$=~4~K with
impurities in the unitary limit.}}
\end{minipage}
\hfill
\begin{minipage}{0.3\textwidth}
\vspace{-4mm}
\includegraphics[keepaspectratio,width=\textwidth]{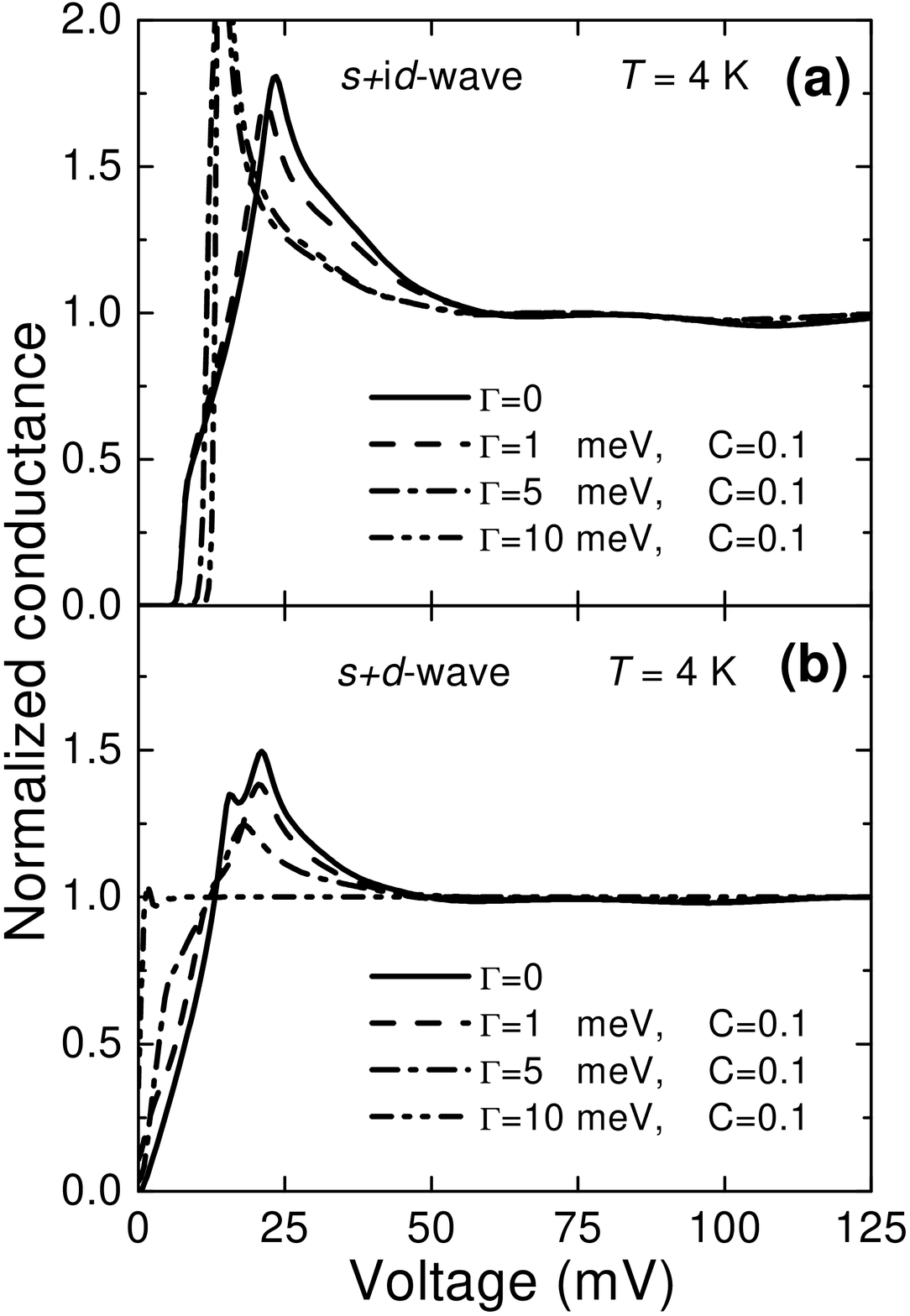}
\caption{{\small Theoretical normalized conductance curves for
$\mu ^{*}\:$=~0 in the $s+$i$d$ and $s+d$ symmetries at $T$=~4~K
with impurities in the unitary limit.}}
\end{minipage}
\hfill
\begin{minipage}{0.3\textwidth}
\includegraphics[keepaspectratio,width=\textwidth]{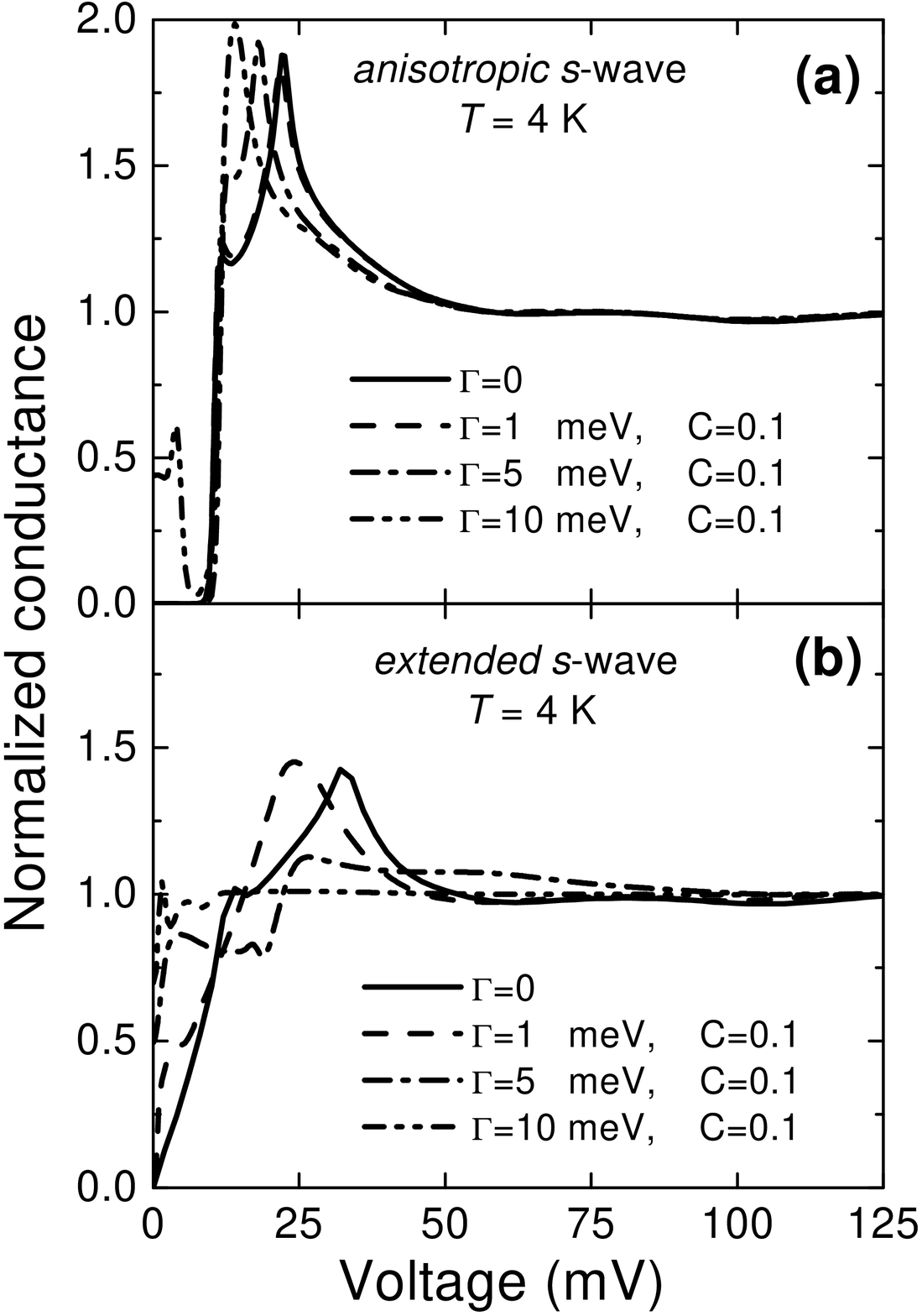}
\caption{{\small Theoretical normalized conductance curves for
$\mu ^{*}\:$=~0 in the {\it anisotropic~s} and {\it extended~s}
symmetries at $T$=~4~K with impurities in the unitary limit.}}
\end{minipage}
\end{figure}
\begin{figure}[b]
\begin{center}\vspace{-6mm}
\includegraphics[keepaspectratio,width=7.3cm]{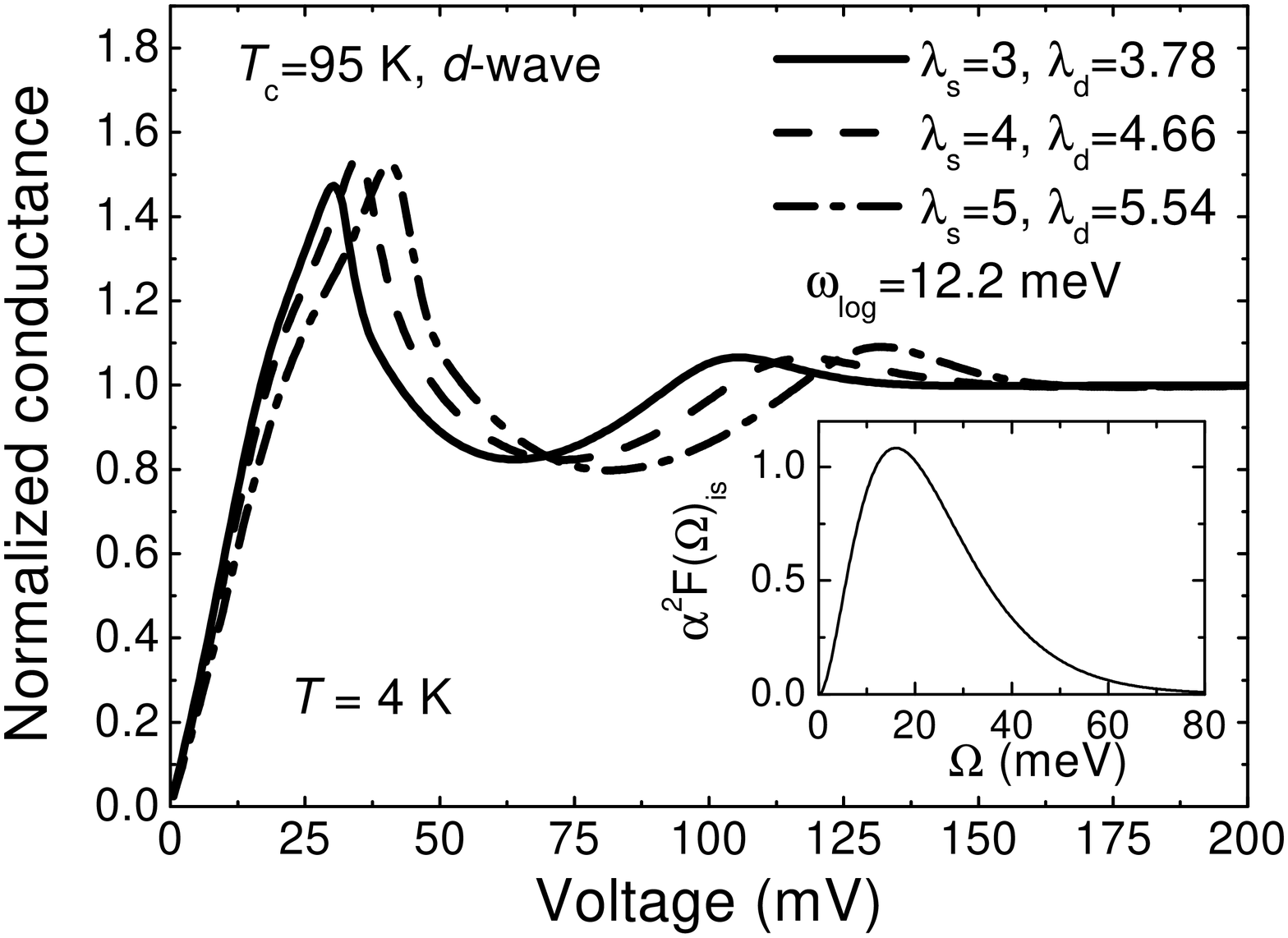}
\vspace{-3mm} \caption{\small{Theoretical $d$-wave conductance
curves calculated at $T=4$~K for increasing values of the coupling
constants. A suppression of the conductivity at
$V\:$=$\:2\Delta\ped{peak}/e$ (`dip') and an enhancement at
$V\:$=$\:3\Delta\ped{peak}/e$ (`hump') appear due to the strong
electron-boson coupling. The inset shows the isotropic
electron-boson spectral function $\alpha^{2}F(\Omega )_{{\rm
is}}$ in the $\lambda\ped{s}=3$ case. In the other cases the curve
is simply scaled along the vertical axis. The anisotropic spectral
function is $\alpha^{2}F(\Omega )_{{\rm an}}=g \cdot
\alpha^{2}F(\Omega )_{{\rm is}}$.}}
\end{center}
\end{figure}

In the unitary limit ($C\ll 1$) the presence of impurities has
very different effects in the various symmetries, as shown in
Figs. 7, 8 and 9 that report the curves obtained with $C=$~0.1 and
$\Gamma $~=~1, 5, 10 meV. Also in this case the results indicate
that, as expected, the main effect of non-magnetic impurities is
to suppress the $d$-wave contribution. In the $d$-wave case, a
small amount of impurities makes the peaks of the tunneling
conductance curves to be lowered, broadened and shifted to the
left, and for $\Gamma \geq 10$~meV the superconductivity
disappears. In the same limit, the $s+$i$d$ and the
\emph{anisotropic}~$s$ curves become very similar to a pure
$s$-wave, but the peak is shifted toward lower voltages. Finally,
the peak of the \emph{extended }$s$ and $s$+$d$ curves is lowered
and broadened and the superconductivity is easily suppressed.

An interesting feature of the theoretical $G\ped{n}(V)$ curves
that occurs for suitable values of the parameters is a
suppression of the tunneling conductance at about twice the gap
value (usually referred to as the `dip') and a consequent
enhancement of conductance which occurs at about three times the
gap value (`hump'). This feature, which has probably its origin
in the non-linear character of the EEs, can be theoretically
obtained at least in two ways. In a previous paper
\cite{bandafinita} we showed that in $s$- and $d$-wave symmetries
it can result from the finiteness of the energy band. In
Ref.\cite{Varelogiannis}, instead, `dip' and `hump' are obtained
in the $s$-wave case by using a very large electron-boson
coupling constant.  Here we extend this result to the $d$-wave
symmetry. Actually, in the $\mu^{\ast}=0$ case large values of
the coupling constants would increase the critical temperature
well beyond its physical value. To conciliate a very strong
coupling with the correct $T_{\mathrm{c}}$, the maximum boson
energy must be reduced, that is, the electron-boson spectral
function must be confined in the low-energy region and,
consequently, $\omega_{\mathrm{log}}$ must be kept sufficiently
small. Greater (and more plausible) values of
$\omega_{\mathrm{log}}$ can be obtained if one also takes
$\mu^{\ast}\neq 0$.

In Fig.~10 we report the theoretical $d$-wave conductance curves
obtained with $\mu^{\ast}=0.2$ and $\omega\ped{c}=500$~meV for
increasing values of the isotropic coupling constant:
$\lambda\ped{is}=3,4,5$. The corresponding anisotropic coupling
constants ($\lambda\ped{an}=3.78,4.66,5.54$) have been chosen so
as to have $T_{\mathrm{c}}^{\mathrm{calc}}\:$=~95~K. The
electron-boson spectral function has the form $\alpha^{2}F(\Omega
)_{{\rm is}}= b \Omega^{2}exp(-\Omega/c)$ where $c$~=8~meV and
$b$~=0.0234, 0.0313, 0.0391~meV$^{-2}$ for
$\lambda\ped{is}=3,4,5$. In all cases $\omega\ped{log}=12.2$~meV,
which is consistent with the energy of the first peak in the
phonon spectral function of HTSC. It is clearly seen that the
conductance curves reported in the figure show a `dip' and a
`hump' at approximately $2\Delta\ped{peak}/e$ and
$3\Delta\ped{peak}/e$.

Features very similar to those discussed so far have been indeed
recently observed in the experimental conductance curves of some
HTSC and reported in literature~\cite{dip}. Notice that some
possible explanations have already been proposed for these
effects \cite{Abanov, Combescot}.

\section{An illustrative application}

Up to now we have dealt only with theoretical predictions, given
by the Migdal-Eliashberg theory, concerning the conductance
curves corresponding to different gap symmetries. We have
discussed the effect of the temperature and of the scattering by
impurities.

Now we can give an example of how these calculations can be
applied to a real physical problem. Actually, the Eliashberg
theory has been successfully used to describe most of the
low-$T\ped{c}$ superconductors by using a simple $s$-wave gap
symmetry, thus it is not worthwhile to use these compounds to test
our model.  Moreover, we have performed our calculations in the
2-dimensional case, and therefore we should refer to layered
materials. We can attempt a comparison of our theoretical
predictions to tunneling conductance curves of SIN junctions
involving high-$T\ped{c}$ superconductors (HTSC). Of course in
applying this theory to HTSC, we must take into account that: i)
$\alpha\ped{an}^2F(\Omega)$ doesn't necessarily have to be
multiple of $\alpha\ped{is}^2F(\Omega)$; ii) $\mu^{\ast}$ is
probably different from zero; iii) as well known, the Fermi
surface (FS) of these materials is not a circle; iv) the bandwidth
is finite; v) the conductance of the junctions generally depends
on geometrical factors which have been disregarded here. Actually,
both $\alpha^2F(\Omega,\phi,\phi^{\prime})$ and $\mu^{\ast}$ are
unknown, and our assumptions are the simplest possible. The
hypothesis that the FS is a circle is as well an
oversimplification, but it allows obtaining a more handy model. As
far as the bandwidth is concerned, we can presume on the basis of
recent results that in optimally-doped HTSC
$\varepsilon\ped{F}\gtrsim400$~meV and therefore the effects of
the finiteness of the bandwidth are not so important
\cite{bandafinita}.

In general, the very complex phenomenology of HTSC yields to the
conclusion that many features of these compounds are probably
beyond the limits of the Eliashberg theory. Then, we don't expect
this theory to provide an explanation for high-$T\ped{c}$
superconductivity. We will simply try to use our model to obtain,
with suitable choices of the parameters, theoretical curves in
agreement with experimental tunneling data in optimally-doped
Bi$_2$Sr$_2$CaCu$_2$O$_{8+\delta}$ (BSCCO)~\cite{PhysC97},
YBa$_2$Cu$_3$O$_{7-\delta}$ (YBCO)~\cite{YBCO},
Tl$_2$Ba$_2$CuO$_{6+\delta}$ (TBCO)~\cite{TBCO} and
HgBa$_2$CuO$_{4+\delta}$ (HBCO)~\cite{HBCO} single crystals. Of
course, the spectral function $\alpha ^{2}F(\Omega ,\phi ,\phi%
^{\prime })$ is unknown. As already pointed out, in the case of
BSCCO we can take, according
to eq.(\ref{alfaBi}): 
\[\alpha ^{2}F(\Omega
)_{{\rm is}}=\frac{\lambda _{{\rm is}}}{\lambda }\alpha
^{2}F(\Omega )_{{\rm Bi2212}}\] \cite{PhysC97} and for the other
materials \vspace{-3mm}
\begin{equation}
\alpha ^{2}F(\Omega )_{{\rm is}}=\frac{\lambda _{{\rm
is}}}{\lambda }G(\Omega ) \label{alfa2n}
\end{equation}
\noindent where $G(\Omega )$ is the phonon spectral density
determined by neutron scattering \cite{neutron}, $\lambda$ is the
corresponding coupling constant and $\lambda_{\mathrm{is}}$ is a
free scaling parameter which must be adjusted to fit the
experimental data. According to our previous assumptions, the
anisotropic component of the spectral function is given by
$\alpha ^{2}F(\Omega)_{{\rm an}}$=$g\cdot \alpha^{2}F
(\Omega)_{{\rm is}}$, where the constant $g$ is another free
parameter. Since in most of the experimental curves a large
zero-bias conductance is present, we will solve the EEs
(\ref{EE1})-(\ref{EE3}) including the term which describes the
scattering from impurities.

Before going on with the discussion, let us focus for a while on
the experimental data reported for BSCCO. As can be observed in
Fig.~10, the break-junction tunneling measurements reported here
give a conductance peak at an energy $\Delta\ped{peak}\equiv
eV\ped{peak}=25$~meV, in good agreement with other break-junction
data appeared in literature \cite{Hancotte}. However, this value
is much smaller than that measured by other techniques, namely the
scanning tunneling spectroscopy, which gives
$\Delta\ped{peak}=37\div 43$~meV \cite{DeWilde,Renner,Davis}. The
reason for these discrepancies is not clear. On the other hand,
we have already shown \cite{PhysC99} that a large gap value such
that measured by STM cannot be reconciled, in the framework of
the Eliashberg theory, with the observed $T\ped{c}$. On the
contrary, the gap obtained from our data can be well reproduced
by our model. For this reason, in the following we will choose to
analyze our break-junction experimental data on BSCCO.

We can now discuss the results of the comparison of experimental
conductance curves to the theoretical ones calculated in the $d$
and $s$+i$d$ symmetries. The choice of these symmetries is
motivated by many experimental evidences for the existence of a
$d$-wave component of the order parameter in most of the
superconducting copper-oxides. The agreement of theoretical and
experimental curves is surprisingly good despite all the
limitation cited above and the roughness of the model.  In
particular, the theoretical curves which give best results are
those obtained in the $s$+i$d$ symmetry in the cases of TBCO and
HBCO, and in the $d$-wave symmetry in the other cases.
Incidentally, it is worthwhile to stress that a good agreement can
be obtained for different values of $\lambda_{\mathrm{is}}$ and
$g$ and, therefore, what we present here is only one of the
possible choices.

The results of the comparison are shown in Fig.~11, and the values
of the parameters are reported in Table~II.

\begin{table}[b]
\begin{center}
\def \V {\rule{0pt}{2.5ex}}
\def \D {\rule[-1ex]{0pt}{0pt}}
\begin{tabular}[h]{|c c|c|c|c|c c|}
\hline \V & & BSCCO & YBCO & TBCO & HBCO & \D \\ \hline \hline \V
& $\lambda\ped{is}$ & 2 & 3.2 & 2 & 3 & \D
\\ \hline \V & $\lambda\ped{an}$ & 2.32 & 3.12 & 2.2 & 3.6 & \D
\\ \hline \V & $\Gamma$~(meV) & 0.5 & 0.6 & 0 & 0 & \D
\\ \hline \V & $C$ & 0.1 & 0.1 & - & - & \D
\\ \hline \V & $T\ped{c}^{\mathrm{exp}}$~(K) & 93 & 90 & 91 & 97 & \D
\\ \hline \V & $T\ped{c}^{\mathrm{calc}}$~(K) & 93 & 90 & 91 & 143 & \D
\\ \hline
\end{tabular}
\end{center}
\caption{Values of the parameters used in the calculations to
obtain the theoretical normalized conductance curves that are
compared to the experimental data for different high-$T\ped{c}$
superconductors.}
\end{table}

\begin{figure}[t]\vspace{-7mm}
\begin{center}
\includegraphics[keepaspectratio,width=0.8\textwidth]{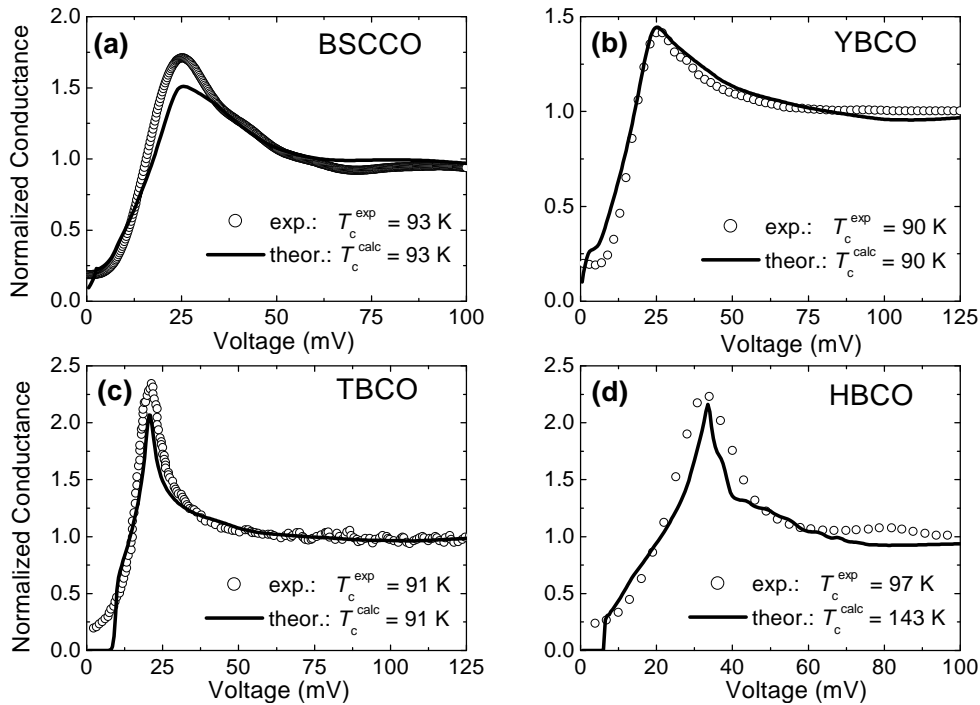}
\end{center}\vspace{-7mm}
\caption{\small{Comparison of the theoretical normalized tunneling
conductance curves to the experimental data obtained on BSCCO~(a)
\cite{PhysC97}, YBCO~(b) \cite{YBCO}, TBCO~(c) \cite{TBCO},
HBCO~(d) \cite{HBCO}. The symmetry of the order parameter used in
the calculations is $d$-wave in cases (a) and (b), and
$s$+i$d$-wave in cases (c) and (d). For details on the values of
the parameters see Table II.}} \vspace{-2mm}
\end{figure}

In the cases of BSCCO and YBCO the theoretical curves follow well
the behaviour of experimental data, and also the calculated
$T_{\mathrm{c}}$ agrees with the measured one. In both these
cases, due to the low zero-bias anomaly ($\approx\,$0.2), the best
result is obtained with a very small impurity content ($\Gamma =
0.5$~meV and $\Gamma=0.6$~meV, respectively). The zero-bias has
approximately the same value also in the cases of TBCO (0.19) and
HBCO (0.23). However, in these cases the best agreement is
obtained with $\Gamma$~=~0, because in the presence of impurities
the peak of the theoretical curve results too low with respect to
the experimental one. It is easily seen that the agreement is
better for TBCO than for HBCO. In this latter case, to obtain an
acceptable result we were forced to use parameters that give a
calculated critical temperature
$T_{\mathrm{c}}^{\mathrm{calc}}=143$~K which is much greater than
the measured one ($T_{\mathrm{c}}^{\mathrm{exp}}=97$~K), while in
all the other cases
$T_{\mathrm{c}}^{\mathrm{calc}}\approx%
T_{\mathrm{c}}^{\mathrm{exp}}$.

Let us conclude with a brief discussion of the assumption
$\mu^{\ast}=0$ we made at the beginning of this paper. This
simplifying assumption could appear poorly adequate to describe
the HTSC, where the Coulomb pseudopotential is, as well known,
different from zero. Actually, the main effect of each component
of $\mu^{*}$ (isotropic and anisotropic) is to change the scale of
the corresponding coupling strength. Then, almost the same
tunneling curves can be obtained with $\mu^{*}=0$ and given
values of the coupling constants, let's say
$\lambda\ped{is}^{(1)}$ and $\lambda\ped{an}^{(1)}$, or with
$\mu^{*}\ped{is}$ and $\mu^{*}\ped{an}$ different from zero and
coupling constants $\lambda\ped{is}^{(2)}>\lambda\ped{is}^{(1)}$
and $\lambda\ped{an}^{(2)}>\lambda\ped{an}^{(1)}$. \vspace{-2mm}

\section{Conclusions}\vspace{-2mm}
We calculated the theoretical normalized tunneling conductance of
SIN junctions for six different symmetries of the superconducting
order parameter, by solving the real-axis EEs both in the
presence and in the absence of scattering from non-magnetic
impurities. We demonstrated that, in spite of being a much more
complicated and time-consuming procedure, this approach is
preferable to the analytical continuation of the solution of the
imaginary-axis EEs, especially if a comparison of the theoretical
predictions with good experimental data is required.

For each of the symmetries considered, we also studied the
temperature dependence of the tunneling conductance curves,
showing that at the increase of temperature the information about
the pairing symmetry is progressively lost.

Finally, we discussed an example of how our calculations can be
applied to a physical system, by comparing the theoretical curves
to experimental tunneling data obtained on BSCCO, YBCO, TBCO and
HBCO. Of course, it must be borne in mind that the applicability
of the Eliashberg theory to these materials is controversial, and
the reliability of the results is limited by the very complex
phenomenology of high-$T\ped{c}$ superconductors.

Unexpectedly, the results of such a comparison are fairly good.
The theoretical curves reproduce some of the unusual properties of
the tunneling curves of HTSC: the high $T\ped{c}$, the broadening
of the conductance peak, the conductance excess below the gap, the
zero bias and also the presence of a `dip' and a `hump' at
2$\Delta/e$ and $3\Delta/e$, respectively. The results also seem
to confirm, for these materials, the plausibility of a $d$-wave
gap symmetry, or at least of a dominant $d$-wave component of the
order parameter.

Even though alternative approaches to high-$T\ped{c}$
superconductivity are nowadays preferred, maybe all these results
indicate that, as far as the tunneling conductance is concerned,
the Eliashberg theory mimics a (still unknown) theory of
high-$T\ped{c}$ superconductivity, whose mechanism could at least
partially differ from the electron-boson coupling.

\end{document}